\documentclass[aps,prl,twocolumn,superscriptaddress,nopacs]{revtex4}

\usepackage{graphicx}
\usepackage{dcolumn} 
\usepackage{bm}      

\begin{document}

\title{Nanoscale freezing of the 2D spin liquid Pr$_{3}$Ga$_{5}$SiO$_{14}$}

\author{H.~D.~Zhou}
\affiliation{Department of Physics, Florida State University,
Tallahassee, FL 32306-3016, USA} \affiliation{National High Magnetic
Field Laboratory, Florida State University, Tallahassee, FL
32306-4005, USA}

\author{C.~R.~Wiebe}
\email{cwiebe@magnet.fsu.edu} \affiliation{Department of Physics,
Florida State University, Tallahassee, FL 32306-3016, USA}
\affiliation{National High Magnetic Field Laboratory, Florida State
University, Tallahassee, FL 32306-4005, USA}

\author{Y.~J.~Yo}
\affiliation{National High Magnetic Field Laboratory, Florida State
University, Tallahassee, FL 32306-4005, USA}

\author{L.~Balicas}
\affiliation{National High Magnetic Field Laboratory, Florida State
University, Tallahassee, FL 32306-4005, USA}

\author{Y.~Takano}
\affiliation{Department of Physics, University of Florida, Gainesville, Florida 32611, USA}

\author{M.~J.~Case}
\affiliation{National High Magnetic Field Laboratory, Florida State
University, Tallahassee, FL 32306-4005, USA}

\author{Y.~Qiu}
\affiliation{NIST Center for Neutron Research, Gaithersburg,
Maryland, 20899-5682, USA} \affiliation{Department of Materials
Science and Engineering, University of Maryland, College Park,
Maryland, 20742, USA }

\author{J.~R.~D.~Copley}
\affiliation{NIST Center for Neutron Research, Gaithersburg,
Maryland, 20899-8102, USA}

\author{J.~S.~Gardner}
\affiliation{NIST Center for Neutron Research, Gaithersburg,
Maryland, 20899-5682, USA} \affiliation{Indiana University, 2401
Milo B. Sampson Lane, Bloomington, IN 47408, USA }

\date{\today}

\begin{abstract}
In this letter, we report
on the single crystal growth and physical characterization of the distorted kagom\'{e} system Pr$_3$Ga$_5$SiO$_{14}$.  It is found that at zero magnetic field the system shows no
magnetic order down to 0.035 K and exhibits a $T^{2}$ behavior for the specific heat at low temperatures, indicative of a gapless 2D spin liquid state.  Application of an applied field induces nanoscale islands of ordered spins, with a concomitant reduction of the $T^{2}$ specific heat term.  This state could be a possible ferro-spin nematic ordering stabilized out of an unusual spin liquid state.
\end{abstract}

\maketitle

One of the desired targets of modern materials science is the realization of model magnetic systems to test physical laws.  In the field of geometrically frustrated magnetism, the Hamiltonians for many lattices have been expressed for various two dimensional and three dimensional sublattice, such as the triangular lattice and the pyrochlores, and experimental analogues have been well studied\cite{Greedan}.  However, the two dimensional kagom\'{e} lattice has proved to be more difficult to synthesize, as many examples found in the literature (such as SrGa$_{8-x}$Ga$_{4+x}$O$_{19}$, or SCGO) have site disorder issues, or are difficult to produce in single crystalline form\cite{Ramirez}.  Jarosites, Such as (D$_3$O)Fe$_3$(SO$_4$)$_2$(OD)$_6$, have a similar problem with a significant amount of Fe vacancies for many species of this family\cite{Wills1}.  Recently, the paratacamite family Zn$_{x}$Cu$_{4-x}$(OH)$_6$Cl$_{12}$\cite{Helton}, which shows a spin liquid state, was
reported. These materials not only have ``structurally perfect'' bond angles for the kagom\'{e} lattice - they are also composed of s = 1/2 Cu$^{2+}$ spins.  It has been shown that these materials are excellent testing grounds for the resonating-valence-bond state predicted by Anderson\cite{Lee}, which has relevance not only for the study of quantum spin liquids, but also for studies of high-temperature superconductors and other low-dimensional magnetic systems. However, the issue of site disorder of the paratacamite still needs to be addressed.

 Many quantum and classical theories for the kagom\'{e} lattice predict ordered ground states (typically with a $\sqrt{3}$ $\times$ $\sqrt{3}$ structure).  The exceptions to this trend include SCGO, the jarosite (D$_3$O)Fe$_3$(SO$_4$)$_2$(OD)$_6$, and the newly discovered paratacamites Zn$_{x}$Cu$_{4-x}$(OH)$_6$Cl$_{12}$.  All of these compounds exhibit unconventional spin freezing at low temperatures.  The former compounds exhibit, for example, a $T^2$ magnetic component to the specific heat (rather than linear component as expected for a spin glass)\cite{Ramirez}.  The paratacamite shows a sublinear $T$ component that is believed to be the result of a Fermi surface of quantum excitations appearing at low temperatures, although this has yet to be confirmed with single crystals\cite{Hermele}.  All of these samples show dynamics on slow time scales that can be resolved with high resolution inelastic neutron scattering\cite{Gaulin}, muon spin relaxation\cite{Uemura} or NMR\cite{Olariu}.  The origin for these slow spin dynamics is thought to lie with persistent two dimensional quantum fluctuations which stabilize a spin liquid state, even in the case of the larger spins with Cr$^{3+}$ (S=3/2) and Fe$^{3+}$ (S = 5/2).

In this letter, we report the single crystal growth and characterization of the distorted kagom\'{e} system Pr$_3$Ga$_5$SiO$_{14}$.  This material has a two dimensional lattice of Pr$^{3+}$ spins (J = 4) with weak antiferromagnetic coupling ($\theta_{\text{CW}}$ = $-$2.3 K).  Neutron scattering experiments fail to show any ordering in zero applied magnetic fields down to 35 mK, despite the presence of a broad peak in the heat capacity at $T$ = 6.7 K.  The presence of a $T^2$ component in the heat capacity is correlated with low energy two dimensional spin excitations.  Application of an external magnetic field perpendicular to the two dimensional layers induces a gap in the spin excitation spectrum, but magnetic Bragg peaks indicative of ordering do not appear.  There is only diffuse scattering present up to H = 9 T which is modeled with the formation of nanoscale islands of ordered spins.  This sort of ordering has been seen in other two dimensional systems such as NiGa$_2$S$_4$\cite{Nakatsuji}, but has only now been observed in a kagom\'{e} lattice with applied fields.  This state could be a possible ferro-spin nematic ordering stabilized out of an unusual spin liquid state.

A single crystal of Pr$_{3}$Ga$_{5}$SiO$_{14}$ was grown by the
traveling-solvent floating-zone technique.  The
room-temperature structure was determined with an X-ray
diffractometer equipped with Cu K$_{\alpha1}$ radiation. X-ray Laue
diffraction was used to orient the crystal. The
magnetic-susceptibility measurements were made with a DC
superconducting interference device (SQUID) magnetometer; the
measurements were made on heating after cooling in zero field
and with applied magnetic field (H = 1 T) parallel to the
longest dimension of the sample.  The specific heat and AC
susceptibility measurements were made with a physical property
measurement system on single crystals with applied fields parallel to
$c$ axis. A dilution fridge was used to collect specific heat data below 0.4 K.
Neutron scattering measurements were completed at the NIST
CHRNS using the Disk Chopper Spectrometer with a wavelength of
5.0 {\AA}. The crystal (total mass of 5 g) was aligned in the $ab$
plane with a vertical magnetic field applied in the $c$ direction. A
dilution fridge was used which had a base temperature of 0.035 K

\begin{figure}[tbp]
\linespread{1}
\par
\begin{center}
\includegraphics[width=90mm,height=70mm,angle=0]{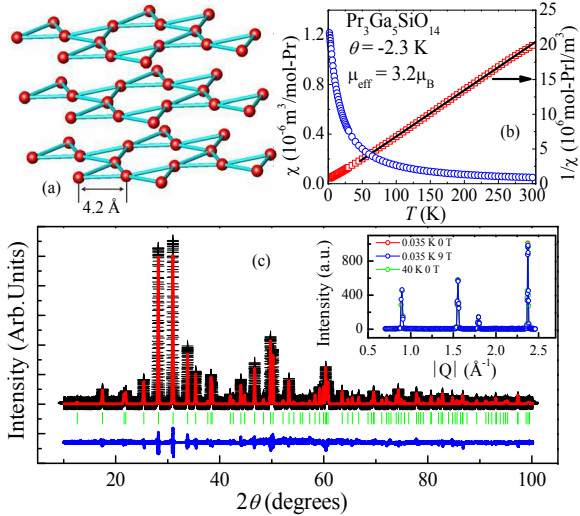}
\end{center}
\par
\caption{(a) Distorted kagom\'{e} lattice of Pr$^{3+}$
ions in the $ab$ plane of Pr$_{3}$Ga$_{5}$SiO$_{14}$. (b)
Temperature dependences of DC susceptibility and its inverse. (c)
XRD pattern for Pr$_{3}$Ga$_{5}$SiO$_{14}$ (plus marks) at room
temperature. The solid curve is the best fit from the Rietveld
refinement using FullProf. The vertical marks indicate the position
of Bragg peaks, and the bottom curve shows the difference between
the observed and calculated intensities. Inset: Neutron diffraction
patterns at $T$ = 0.035 K, H = 0 T; $T$ = 0.035 K, H = 9 T; and $T$
= 40 K, H = 0 T show no magnetic Bragg peaks.}
\end{figure}

Pr$_{3}$Ga$_{5}$SiO$_{14}$ crystallizes in the
trigonal space group \emph{P}321 with lattice parameters \emph{a} =
8.0661(2) {\AA} and \emph{c} = 5.0620(2) {\AA} (Fig. 1(c)). The
Pr$^{3+}$ magnetic ions in Pr$_{3}$Ga$_{5}$SiO$_{14}$ are organized
in corner sharing triangles in well-separated planes perpendicular
to the \emph{c} axis. Within each plane, the Pr$^{3+}$ ions form a
distorted kagom\'{e} lattice (Fig. 1(a)), which is topologically
equivalent to the ideal kagom\'{e} when only the shortest atom
bridging interactions are considered. The nearest Pr-Pr separation in the
\emph{ab} plane is 4.2 {\AA} (Fig. 1(a)).

Figure 1(b) shows the temperature dependences of the DC magnetic
susceptibility ($\chi(T)$) and its inverse ($\chi^{-1}(T)$).
The susceptibility above 50 K follows the Curie-Weiss law:
$\chi^{-1}(T)$ = ($T - \theta_{\text{CW}})/C$. The effective moment
$\mu_{\text{eff}}$ = 3.2(1) $\mu_{\text{B}}$ calculated from the
Curie constant \emph{C} is smaller than that of single Pr$^{3+}$ ion
($\mu_{\text{eff}}$ = 3.6 $\mu_{\text{B}}$), which is common for
magnetically frustrated systems\cite{Greedan}. The Curie-Weiss
temperature, $\theta_{\text{CW}}$ = $-$2.3 K, indicates that the nearest neighbor
 interactions are antiferromagnetic. No magnetic anomalies were observed
down to 1.8 K from the DC susceptibility data. The inset of Fig. 1 (c)
shows the elastic neutron diffraction patterns at different
temperatures and magnetic fields. No magnetic Bragg peaks or significant magnetic diffuse scattering features are
found down to 0.035 K. Note that this gives a frustration index of
$f \sim \mid\theta_{\text{CW}}\mid/T_{\text{C}} \geq 66$.

\begin{figure}[tbp]
\linespread{1}
\par
\begin{center}
\includegraphics[width=90mm,height=70mm,angle=0]{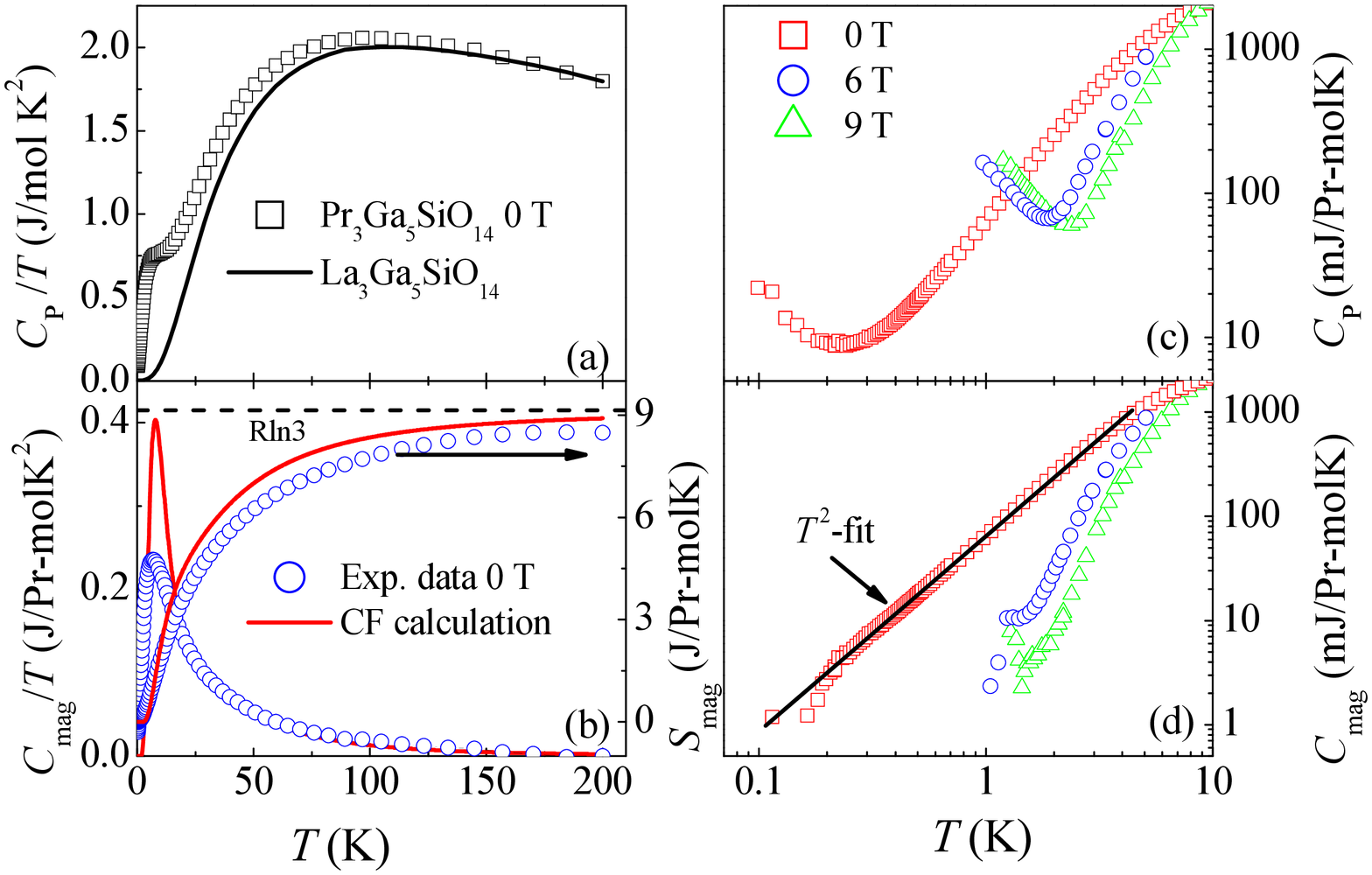}
\end{center}
\par
\caption{(a) Temperature dependences of $C_{\text{P}}/T$
for Pr$_{3}$Ga$_{5}$SiO$_{14}$ and La$_{3}$Ga$_{5}$SiO$_{14}$. (b)
Temperature dependences of $C_{\text{mag}}/T$ and the integrated
entropy for Pr$_{3}$Ga$_{5}$SiO$_{14}$ and the crystal field
calculation. Open circles are experimental results, solid lines are
calculated results. (c) Temperature dependences of the specific heat at
low temperatures for Pr$_{3}$Ga$_{5}$SiO$_{14}$ with H = 0 T, 6 T, and 9 T. (d)
Magnetic contribution of specific heat after subtracting the
Schottky anomaly and lattice contribution.}
\end{figure}

The specific heat data, $C_{\text{P}}(T)$, also shows no evidence of a
phase transition down to 0.1 K in zero field (Fig.~2(a) and (c)). The
magnetic specific heat $C_{\text{mag}}(T)/T$ after substraction of the
lattice contribution ($C_{\text{P}}(T)$ of
La$_{3}$Ga$_{5}$SiO$_{14}$ with no magnetic ions) exhibits a broad
peak at $T_{\text{peak}}$ = 6.7 K. The integration of
$C_{\text{mag}}/T$ gives an entropy $S_{\text{mag}}$ = 8.5 J/mol-K (
circles in Fig.~2(b)) near 200 K, which is approaching $R$ln3 ($R$
is the gas constant).  This value suggests that the crystal field
scheme of Pr$^{3+}$ below 200 K could be (i) three singlet levels,
(ii) a doublet ground state and a singlet excitation level, or (iii)
a singlet ground state and a doublet excitation level. These three
possibilities have been tested with the experimental data - the best
fit is obtained from the three singlet level scheme with $\Delta_{1}$ =
25 K and $\Delta_{2}$ = 117 K. This is in good agreement with
earlier studies on the related garnet structure
Pr$_{3}$Ga$_{5}$O$_{12}$, which has three singlets with an energy
separation of $\Delta_{1}$ = 26 K and $\Delta_{2}$ = 68 K\cite{Hooge}. It would
be unusual if Pr$_{3}$Ga$_{5}$SiO$_{14}$, of lower site symmetry,
would have doublets if the cubic garnet structure is already composed of
low temperature singlets. The next singlet of
Pr$_{3}$Ga$_{5}$O$_{12}$ is at $T$ = 780 K, and this is not seen in our
specific heat data. The calculated data (the line in Fig.~2(b)) fits
experimental data well at high temperatures; but at low temperatures
the experimental data shows a more broad peak and larger value of
specific heat below $T_{\text{peak}}$. The origin for this feature
is the broadening of the crystal field levels due to correlation
effects, as seen in other spin liquid candidates such as
Tb$_2$Ti$_2$O$_7$\cite{Gingras}.

At low temperatures (Fig.~2(c)), the nuclear Schottky anomaly introduces an upturn in the specific heat,
which can be fit with a $T^{-2}$ term.  After the substraction of this anomaly and the lattice contribution,
the magnetic contribution of the specific heat at zero field exhibits power-law behavior at low temperatures (Fig.
2(d)). The data between 0.1 K and 4 K are well fitted by a power
law $C_{\text{mag}} = AT^{\alpha}$, where $A$ is a constant and
$\alpha$ = 1.98(2). Two dimensional (2D) spin excitations would give $C
\sim T^{2}$.  This quadratic temperature dependence without
long-range magnetic order indicates the presence of gapless
linear modes in 2D, similar to other 2D kagom\'{e} systems such as SCGO.\cite{Ramirez}

In general, a peak in $C(T)/T$ results from a peak in the density of
states, $g(w)$, defined as $U$ = $\int$ $dwg(w)n(w)w$, where $U$ is
the internal energy, $n(w)$ is the Bose population factor, and the
integral is taken over the excitation bandwidth. Usually, the
temperature of the $C(T)/T$ peak is roughly half the mode energy.
Applying this rule here, a peak in $g(w)$ is expected at $\hbar$$w_{0}
\sim$ 13 K. In order to confirm this prediction, inelastic neutron
scattering experiments were completed. In Fig. 3(c), the results of
the integrated intensity scans are shown as a function of applied
field. Note that at zero field there is a broad peak at E = 1.2 meV
$\sim$ 13 K, which is consistent with the temperature of the specific heat peak.

In summary, the zero field data on single crystalline
Pr$_{3}$Ga$_{5}$SiO$_{14}$ shows the following low temperature properties:
(i) the absence of long-range magnetic order; (ii) the absence of
magnetic diffuse scattering; (iii) a $T^{2}$ dependence of the
specific heat; and (iv) the existence of spin excitations related to
a highly degenerate state as $T$ approaches zero. These properties
place strong constraints on possible ground states. Observations
(i), (iii), and (iv) are consistent with a spin liquid with no
conventional long-range magnetic order, such as observed in the
triangular lattice NiGa$_{2}$S$_{4}$\cite{Nakatsuji} and
hyper-kagom\'{e} Na$_{4}$Ir$_{3}$O$_{8}$\cite{Okamoto}. However, the
absence of magnetic diffuse scattering is unusual. One possibility is that the spins are truly dynamic on the neutron time scale.  The AC susceptibility (Fig. 3(a, b)) shows no
anomaly, nor any frequency dependence down to 1.8 K, which also
indicates that the characteristic fluctuation rate is beyond the kHz
region.  The featureless AC susceptibility data also excludes the
possibility that Pr$_{3}$Ga$_{5}$SiO$_{14}$ is a spin glass at low
temperatures.

\begin{figure}[tbp]
\linespread{1}
\par
\begin{center}
\includegraphics[width=75mm,height=90mm,angle=0]{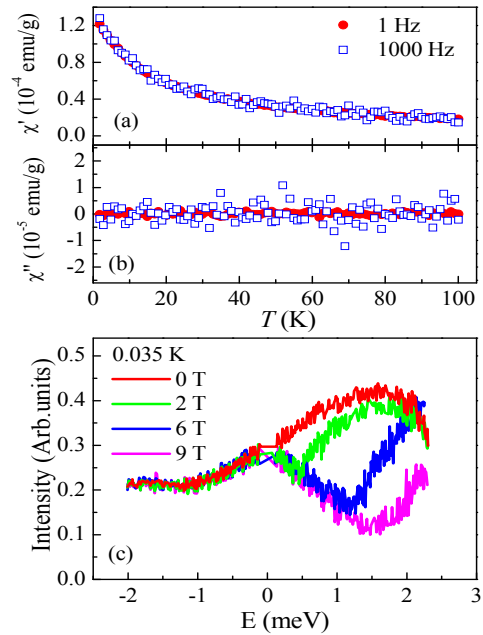}
\end{center}
\par
\caption{Temperature dependences of the AC susceptibility: (a)
real part and (b) imaginary part at frequency 1 Hz and 1000 Hz.
(c) Inelastic neutron scattering integrated over Q at 0.035 K as a function of magnetic field.}
\end{figure}

\begin{figure}[tbp]
\linespread{1}
\par
\begin{center}
\includegraphics[width=75mm,height=90mm,angle=0]{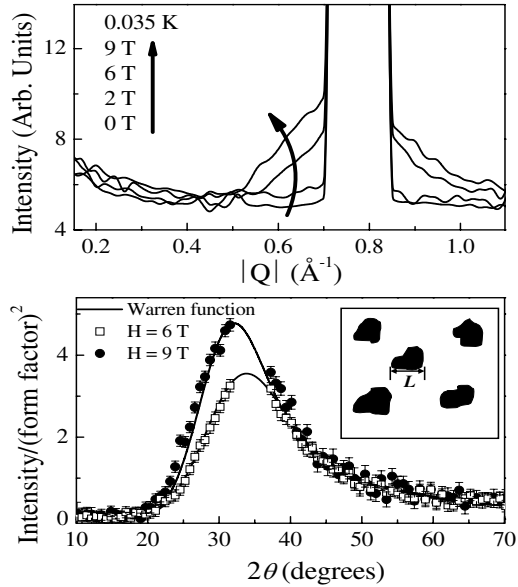}
\end{center}
\par
\caption{(a) Diffuse scattering of
Pr$_{3}$Ga$_{5}$SiO$_{14}$ at 0.035 K with different magnetic
fields. (b) Difference between patterns taken at H = 6 T , 9 T
and that taken at H = 0 T. The solid lines are fits to Eq.~1. Inset: Schematic plot of magnetic clusters with size $L$.}
\end{figure}

The specific heat measured at magnetic fields (Fig.~2(d)) shows
that the absolute value becomes smaller and $T^{2}$
behavior disappears. These field-dependent behaviors can be related to the inelastic spin excitation (Fig.~3(c)), which is
clearly suppressed by applied fields. The suppression of the
density of states of the  2D spin excitations in higher fields is clear from the
neutron scattering data. The elastic neutron pattern at 0.035 K with
H = 9 T (Inset of Fig.~1(c)) shows no extra peak and intensity
compared to the zero field pattern, which means no long-range
magnetic order is stabilized. However, significant diffuse scattering appears near Q = 0.78 {\AA} (Fig.~4(a)) with H = 9 T. The form of this diffuse scattering of a broad asymmetric
peak (a sharp rise at low Q and a slow fall off toward high Q) is
characteristic of two-dimensional short-range order.  The
scattering from Pr$_{3}$Ga$_{5}$SiO$_{14}$ can be described
analytically by a modified Warren function for 2D magnetic correlations\cite{Wills2}. The structure
factor $P_{2\theta}$, around the peak is expressed by:

\begin{equation}
 P_{2\theta} = Km\frac{F_{hk}^{2}(1+\text{cos}^{2}2\theta)}{2(\text{sin}\theta)^{3/2}}(\frac{L}{\sqrt{\pi}\lambda})^{1/2}F(a)\cdot[J(z)]^{2}
\end{equation}

with
\begin{equation}
 a = (2\sqrt{\pi}L/\lambda)(\text{sin}\theta-\text{sin}\theta_{0})
\end{equation}

and with
\begin{equation}
 F(a) = \int^{10}_{0}\text{exp}[-(x^2-a)^2]\,dx
\end{equation}

$L$ is the spin-spin correlation length, $\lambda$ is the
wavelength, $K$ is a scaling constant, $m$ is the multiplicity of
the reflection, $F_{hk}$ is the two-dimensional structure factor for
the spin array, and $\theta_{0}$ is the center of the peak. The fits
of the diffuse scattering to equation (1) (Fig.~4(b)) give a correlation
length $L = 25(1)~{\text{\AA}}$ and $29(1)~{\text{\AA}}$ for H = 6 T
and 9 T, respectively. This correlation length corresponds to $6
\sim 7$ in-plane kagom\'{e} lattice spacings. The increasing $L$
with increasing field shows that the short-range order or the
nanoscale magnetic cluster size expands with increasing fields.  The formation of these clusters induces a spin gap in the excitation spectrum, and subsequently reduces the $T^{2}$ component of the specific heat.

For the studied geometrically frustrated systems, such as
SCGO\cite{Ramirez},
NiGa$_{2}$S$_{4}$\cite{Nakatsuji}, and
Na$_{4}$Ir$_{3}$O$_{8}$\cite{Okamoto}, one common feature is that the $T^{2}$ behavior of
specific heat is independent of the applied magnetic field, suggesting that the
ground state consists of moment free spin clusters. In contrast,
Pr$_{3}$Ga$_{5}$SiO$_{14}$ studied here has a $T^2$ behavior which is
sensitive to magnetic fields with a gap opening in the spin excitation spectrum.  One of the possible states which could correspond to such behavior is the recently predicted spin nematic that should appear in two dimensional triangule-based lattices\cite{Tsunetsugu, Bha}.  However, most of these models in the literature lack terms in the Hamiltonian to express the crystal field effects of the Pr$^{3+}$ sites (they are modeled on Heisenberg like transition metal systems with quenched orbital angular momentum).  Despite this, we can rule out the presence of a non-collinear spin nematic\cite{Tsunetsugu}, since much of the spin excitation spectrum is gapped out within the $ab$ plane.  A ferro-nematic state, as mentioned by Bhattacharjee $et~al.$\cite{Bha}, in which the moment is spatially uniform and the excitations are polarized, is a more likely possibility.  Future experiments are clearly needed to verify or refute this scenario.

\begin{acknowledgments}

This work was made possible by support through the NSF (DMR-0084173
and DMR-0454672), the EIEG pro- gram (FSU) and the auspices of the
state of Florida. The authors are grateful for the local support
staff at the NIST Center for Neutron Research. Data analysis was
completed with DAVE, which can be obtained at
http://www.ncnr.nist.gov/dave/.

\end{acknowledgments}

\end{document}